# scientific reports

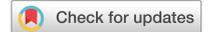

OPEN

# Compact omnidirectional multicore fiber-based vector bending sensor

Josu Amorebieta[1]✉, Angel Ortega-Gomez[1], Gaizka Durana[1], Rubén Fernández[1], Enrique Antonio-Lopez[2], Axel Schülzgen[2], Joseba Zubia[1], Rodrigo Amezcua-Correa[2] & Joel Villatoro[1,3]

We propose and demonstrate a compact and simple vector bending sensor capable of distinguishing any direction and amplitude with high accuracy. The sensor consists of a short segment of asymmetric multicore fiber (MCF) fusion spliced to a standard single mode fiber. The reflection spectrum of such a structure shifts and shrinks in specific manners depending on the direction in which the MCF is bent. By monitoring simultaneously wavelength shift and light power variations, the amplitude and bend direction of the MCF can be unmistakably measured in any orientation, from 0° to 360°. The bending sensor proposed here is highly sensitive even for small bending angles (below 1°).

The demand for sensors that are capable of monitoring in real-time parameters such as amplitude and direction of bending and/or curvature has remained consistent for years due to their relevance in fields such as structural health monitoring (SHM). For this application, such sensors can provide critical information about the status of the structure, fatigue of materials or prevention of cracks, among other important characteristics. This becomes especially significant for building health monitoring systems[1], where bending sensors are used to monitor the inclination of columns or pillars, or for the aerospace industry[2], where they are used to prevent fractures in turbine blades or to test the tolerance of components. In addition, in recent years, new areas in which vector bending measurement is required have emerged, as biomedicine[3] and biomechanics[4], which have caused vector bending sensors to gain visibility and interest.

For such applications, small size and simplicity are compulsory. In this context, the intrinsic properties of optical fibers make them an excellent candidate as they provide high sensitivity, remote sensing possibility, electromagnetic immunity and small size. However, the capability of discriminating the direction of the bend and its amplitude is also demanded, as in the aforementioned applications, bending may occur in different and unexpected directions. The latter is a challenging feature as, usually, optical fiber-based bending sensors are only sensitive to few directions, which restricts their functionality.

The most spread optical fiber-based bending sensors are based on Bragg[5–8] and long period gratings[9–12]. On the one hand, the first ones are very compact, up to few millimeters. However, generally, they have cross sensitivity and require reference sensors to perform correctly, and have low sensitivity for small angles. On the other hand, the second ones require proper isolation from the surrounding in order to be sensitive just to bending, which may degrade their performance. Bending sensors based on interferometry[13,14] and photonic crystal fibers[15] have been reported as well, although they show high complexity, or low reproducibility and sensitivity. In general, such fiber-based solutions require a sensor for each of the directions that is wanted to be detected and measured, resulting in complex or expensive setups.

With a different approach, in the last years, some MCF-based bending sensors have been reported. Those based on MCF with isolated cores[16,17] require complex setups, whereas those based on MCFs with symmetrically arranged coupled cores[18] are incapable of distinguishing the direction of the bending. Lastly, MCFs with asymmetrically disposed coupled-cores or with cores with different refractive indices have shown great potential as vector bending sensors[19–21]. In fact, MCF-based vector bending sensors have demonstrated to have similar sensitivities[19,21] as those reported with other techniques[15]. However, such solutions require a precise alignment between the bending direction and the orientation of the cores, and/or specific modifications of the physical properties of certain cores in the MCF to operate. Moreover, they are only capable of measuring the amplitude of the bending in certain directions by analyzing the wavelength shift or the intensity variation in the spectrum.

To overcome such limitation and design an omnidirectional optical solution, we report on a simple and compact vector bending sensor that is able to discern any direction and amplitude (defined as the angle, quantified in degrees, in which the fiber has been bent) of the bending. It is based on a short segment (8 mm) of MCF fusion spliced to the distal end of a standard single mode fiber (SMF), see Fig. 1a. The device operates in reflection mode and its spectrum is interrogated in wavelength shift and light power variation simultaneously. In this manner, the amplitude and direction of the bending can be unmistakably measured when the MCF is bent, as the spectrum shrinks and shifts accordingly.

[1]Department of Communications Engineering, University of the Basque Country UPV/EHU, 48013 Bilbao, Spain. [2]CREOL-The College of Optics and Photonics, University of Central Florida, Orlando, FL 162700, USA. [3]Ikerbasque-Basque Foundation for Science, 48011 Bilbao, Spain. ✉email: josu.amorebieta@ehu.eus

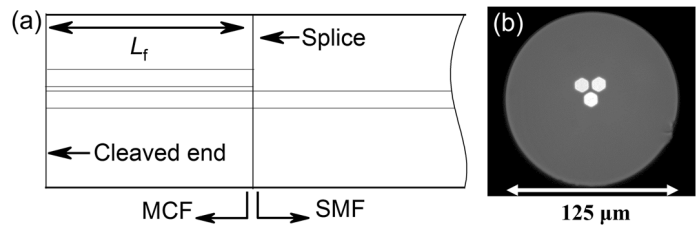

**Figure 1.** (**a**) Schematic representation of the longitudinal arrangement of the sensor (SMF–MCF structure) drawn with Origin2019b (https://www.originlab.com/). (**b**) Cross section of the asymmetric MCF used to build the device.

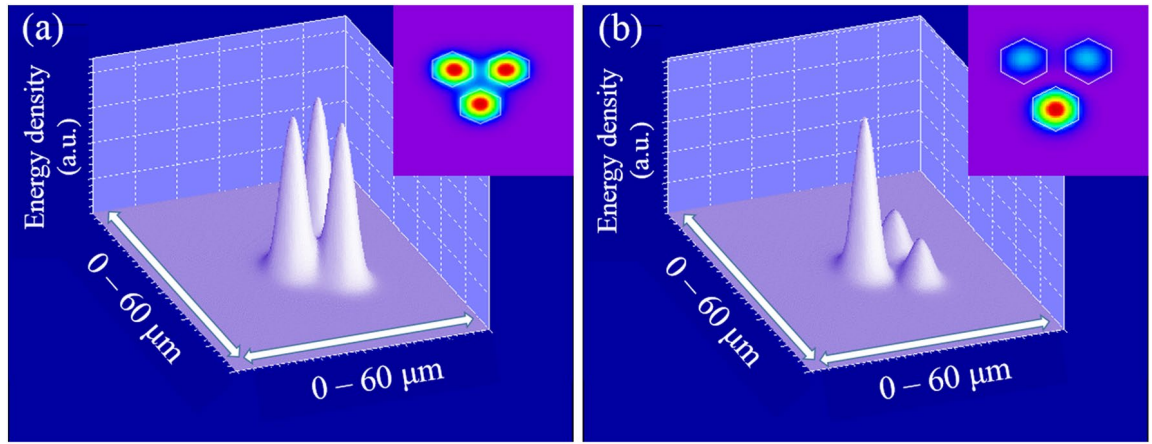

**Figure 2.** Simulations made with PhotonDesign simulation software v6.6 (https://www.photond.com/) of the 3D and 2D profiles of the two coupled orthogonal supermodes excited in the MCF. In (**a**) the supermode $SP_{01}$ is shown, and in (**b**) the supermode $SP_{02}$.

**Sensor design, operating mechanism and fabrication.** The cross section of the MCF used to manufacture the device is shown in Fig. 1b. It consists of three identical and asymmetrically arranged cores made of Ge-doped silica. One of the cores is located at the geometrical center of the MCF, whereas the other two are surrounding it and adjacent to each other. Each core has a mean diameter of 9 µm and has a numerical aperture (NA) of 0.14, identical to that of the SMF (Thorlabs SMF-28-1000). The mean distance from the center of the fiber to the center of neighboring cores is 11.5 µm and they are all inlayed in a pure silica cladding whose outer diameter is 125 µm.

The working mechanism of an SMF–MCF–SMF structure is based on the coupled mode theory (CMT)[22]. According to it, a cyclical power transfer will take place among waveguides if they are close enough to interact. Its particularization for optical fibers and its corresponding theoretical and mathematical analysis is explained in detail in[23,24]. This particularization is valid for strongly coupled MCFs as well, as it is assumed that each of the cores acts as a waveguide. Moreover, the modes propagating through such structures are called supermodes, which are the linear combination of individual $LP_{01}$ modes propagating through each of the cores. As in conventional CMT it is assumed that the propagating modes under study are orthogonal[22], when a strongly coupled MCF is excited in the central core by the fundamental $LP_{01}$ mode of an SMF, the two orthogonal supermodes that have power in the central core will be coupled. Such supermodes, and therefore, the interaction between them, are specific for each MCF design[25,26]. For the MCF described in Fig. 1, the two coupled orthogonal propagating supermodes are shown in Fig. 2.

Considering this, and based on the mathematical expressions for the normalized power coupling in[23,24], it can be concluded that, for MCFs, the power coupling depends on the launched wavelength, on the length of the MCF segment and on the effective refractive indices of the aforementioned orthogonal propagating supermodes. Thus, for the case of the MCF that has been used in this work and described previously, the particularization of such mathematical equations provides the following expression for the normalized coupled power in the central core:

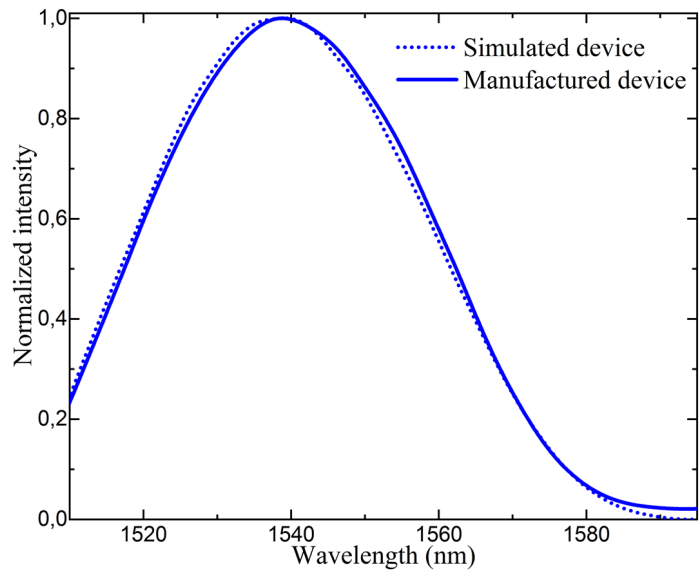

**Figure 3.** Spectra of the manufactured and simulated devices of 8 mm of MCF.

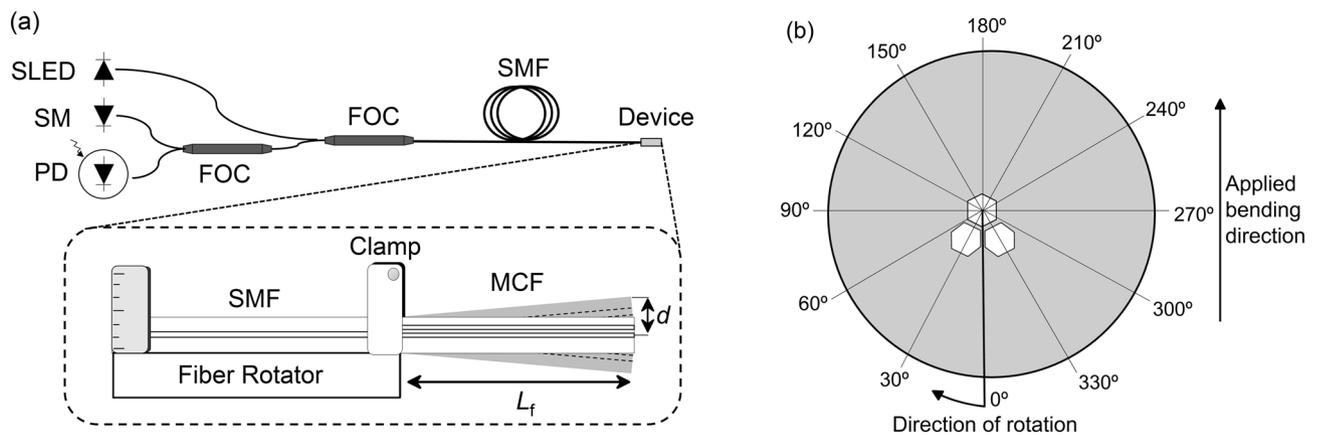

**Figure 4.** (**a**) Schematic of the interrogation setup drawn with Origin2019b (https://www.originlab.com/). The close-up shows how the bending was applied to the sensor. *SLED* superluminescent diode, *SM* spectrometer, *PD* photodetector, *FOC* fiber optic coupler. (**b**) Schematic of the direction of rotation of the MCF, the evaluated points and the applied bending direction drawn with Origin2019b (https://www.originlab.com/).

$$P(z) = \cos^2\left(\frac{\sqrt{3}\pi \Delta n}{\lambda}z\right) + \frac{1}{3}\sin^2\left(\frac{\sqrt{3}\pi \Delta n}{\lambda}z\right) \quad (1)$$

In Eq. (1), $P$ is the coupled power in the central core, $z$ is the distance in the propagation axis at which the normalized coupled power is being calculated, $\lambda$ is the emitted wavelength, and $\Delta n$ is the difference between the effective refractive indices of the two propagating supermodes. In an SMF–MCF–SMF structure, when the light that has propagated through the MCF exits it towards the SMF, only the wavelengths at which there is coupled power in the central core will propagate through the latter. Thus, if light from a broadband source is launched into such structure and it is analyzed with a spectrometer, a series of maxima and minima in the spectra will take place. At a maximum, it indicates that most of the coupled power is at the central core for that wavelength, whereas at a minimum, it indicates that most of the power is coupled to the outer cores for that wavelength.

The sensor proposed in this work has an SMF–MCF structure as shown in Fig. 1a in order to operate in reflection mode, which means that the light is reflected in the cleaved end of the MCF and travels back. Thus, if we suppose an MCF segment of length $L$, it is assumed that the light has propagated along a distance of $2L$ when it exits the MCF towards the SMF. According to that, at that point, the normalized coupled power in the central core for a sensor of length $L$ operating in reflection mode is as expressed in Eq. (1) but particularized for $z = 2L$.

Let us now analyze the case when the MCF is straight or bent. As it has been widely demonstrated[27–29], the refractive index of the core of an optical fiber changes when the fiber is bent. For MCFs, certain core or cores in it are going to be more stressed than others when they are subjected to the same bending plane and amplitude, resulting in sudden changes in the effective refractive indices of the propagating supermodes. According to Eq. (1), this effect implies changes in the normalized power coupling conditions, and therefore, in the resulting spectrum.

Among strongly coupled MCFs, for symmetric ones, the effect of the bending plane will be always the same irrespective of the bending direction. This is why such fibers are not capable to distinguish the direction of bending[18]. However, for asymmetric MCFs, as the one used in this work, the effect of bending in the spectrum will depend on the core orientation and the bending direction as well as on the arrangement and geometry of the cores. As reported on[19–21,30], when these type of MCFs are bent, the variation in the refractive index of each core depends on the bending plane and its orientation with respect to it. As the cores are arranged asymmetrically, each of them suffers different levels of stress against the same bending plane and radius, causing the refractive index of each to vary independently. Such situation affects directly to the propagating supermodes, which modifies consequently the output spectrum of the normalized coupled power in Eq. (1). This is what makes asymmetric MCFs good candidates for direction-sensitive bending sensors, as the variation of the normalized coupled power will be specific for each bending case, giving rise to detectable changes in the spectrum of an SMF–MCF structure in terms of wavelength shift and/or light power variation depending on the applied bending direction.

In order to detect and measure the bending direction causing such changes in the spectrum, by monitoring only one of these two variables, in principle, it could be possible to measure the bending direction and amplitude without ambiguity in 180°. In that range, the sensitivity of the measured variable will be unique for each bending plane and therefore, its measurement (shift to longer and shorter wavelengths, or increase and decrease of the reflected light power) could be linked to the applied bending unequivocally. However, for the remaining 180°, such measurements of the variable are going to be the same as for the previous 180°, giving as a result an ambiguity in which two bending directions provoke the same measured change in the monitored variable.

Such ambiguity can be eliminated with the interrogation system proposed in this work, which consists in simultaneously measuring both variables (the wavelength shift and light power variation), as each of them shows unique sensitivity for each bending direction. In this manner, the measurement of the first variable will provide two possible solutions (two bending directions), whereas the second variable will be the key to resolve such ambiguity. For example, let us assume a case in which a certain wavelength shift has been measured. As it has been explained previously, such shift can be caused by two different bending directions. At this point, the measurement of light power will be the key to resolve the ambiguity, as in one of the possible solutions (one of the bending directions that cause such wavelength shift), the measured light power increases, whereas in the other bending direction, it decreases. This leads to a unique possible bending direction causing such changes in both variables at the same time. Identical procedure can be carried out by inverting the variables: using the light power measurement to provide two possible solutions and the wavelength shift to resolve the ambiguity. Therefore, by combining the simultaneous measurement of both variables, the ambiguity can be resolved and any bending direction can be unmistakably identified in 360°, making the device sensitive to any bending direction (omnidirectional). Such operating principle is demonstrated in this work.

The key to optimize the performance of an omnidirectional MCF bending sensor relies on defining an adequate length ($L_f$) of the MCF. The goal is to obtain a spectrum with one centered and well-defined maximum in the operating window (1510–1595 nm) of the spectrometer, and with no secondary maxima. To fulfill the first design condition, it must be noted the fact that, according to Eq. 1, as the MCF shortens, the period of the normalized coupled power enlarges, as they are inversely proportional. Based on that, the MCF segment should be short in order to have one peak in the interrogation window. According to the second requirement, the aim is to avoid any loss in sensitivity or ambiguity when measuring the power of the reflected light. The need for such requirement is clear if we take into account the results reported in[19], where there are two maxima that have opposite trends in terms of amplitude, which may cause identical power readings for different bending cases. Thus, light power readings would be disabled for such device, limiting it to operate only with wavelength shift, and consequently, limiting notoriously its capabilities as bending sensor, as explained in previous paragraphs.

To obtain an $L_f$ that fulfilled all the aforementioned requirements, PhotonDesign simulation software was used. According to it, the value of $\Delta n$ was $4.8 \times 10^{-4}$ at 1550 nm for this MCF. Simulation results indicated that the best fitting MCF length was 8 mm. This short length makes the sensor very compact, which is a desirable characteristic in optical fiber sensing.

To fabricate the device, a high precision fiber cleaver (Fujikura CT105) was used to obtain an 8-mm-long MCF segment with an accuracy of ± 10 µm. Moreover, in this way, we guaranteed that the cleaved MCF end acted as a low-reflectivity mirror. To obtain the SMF–MCF structure, a precision fusion splicer (Fujikura 100P+) was used, which aligned precisely the central core of the MCF and the core of the SMF. With this machine, and thanks to the matching NAs of the fibers, low insertion losses (below 0.1 dB) can be achieved. The matching between the simulated and manufactured spectra is shown in Fig. 3.

### Results and discussion

The device laid horizontally on a fiber rotator (Thorlabs HFR007) in such a way that the MCF segment was set in a cantilever configuration (see Fig. 4a). This setup allowed selecting accurately the bending plane applied to the MCF. The SMF–MCF fusion splice point of the device was secured by means of the clamp of the fiber rotator, whereas the MCF end was left free, as shown in Fig. 4a. As the cantilevered MCF ($L_f$) was only 8 mm long, it was completely straight in idle position. To set an initial core orientation that would be the reference position from which the fiber would be rotated (the 0° in Fig. 4b), a high-resolution camera (Dyno-Lite AM4116T) was set in front of the MCF to record the orientation of the cores. The initial core orientation that was decided to be used as reference (the aforementioned 0° from which the fiber would be rotated) was when the cores were in an inverted V-like configuration, as indicated in Fig. 4b.

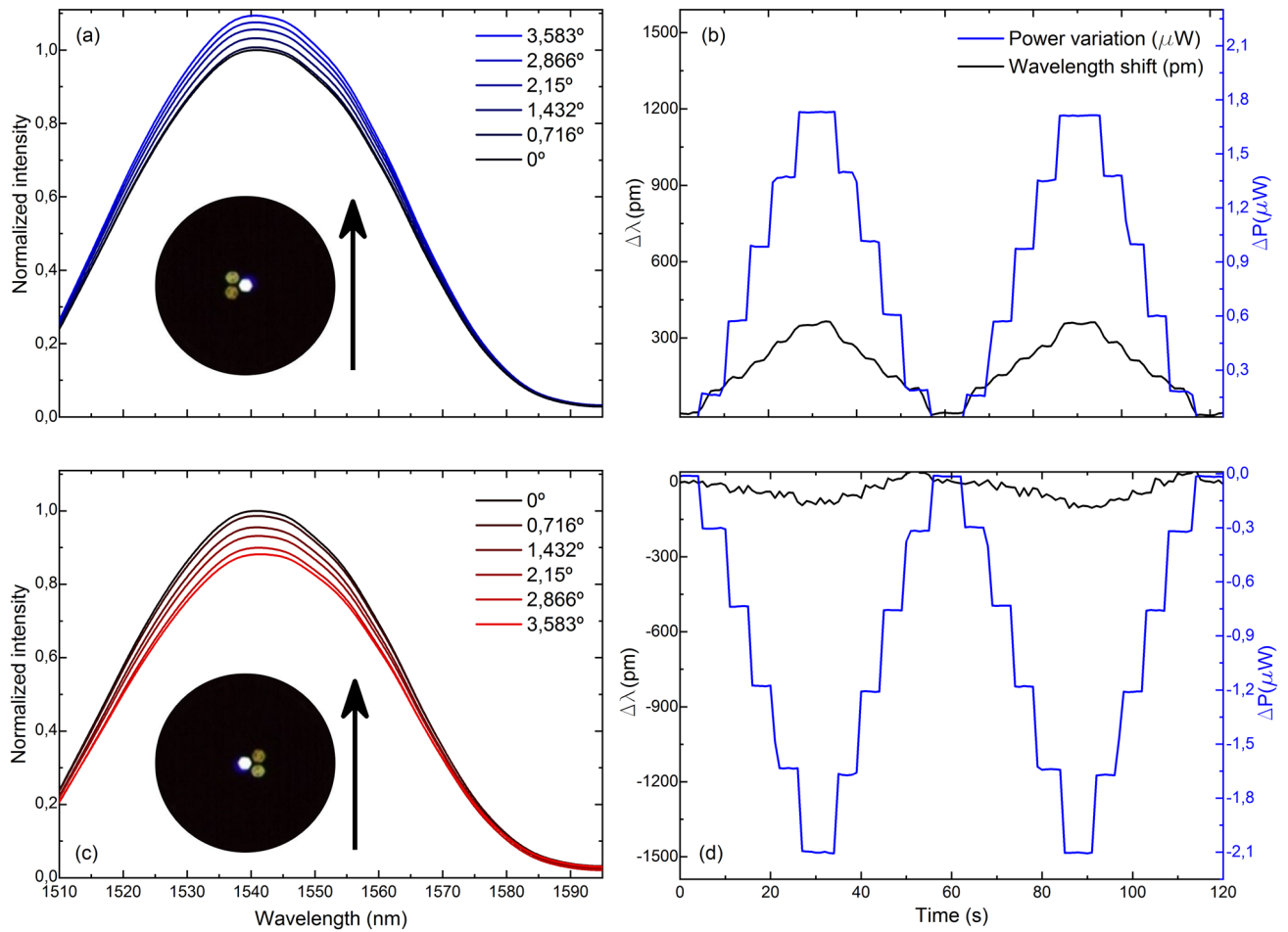

**Figure 5.** Reflection spectra and two cycles of the time evolution of the measured variables when the fabricated device is bent in the direction indicated by the black arrow and the MCF is rotated (**a**, **b**) 90° and (**c**, **d**) 270° with respect to the initial 0° position, respectively.

Once the MCF had its cores in that orientation, a thin ceramic tube (Omega Engineering TRX-005132-6), whose bore's inner diameter (127 μm) was slightly bigger than the diameter of the MCF (125 μm), was set in front of it, and by means of a micrometric displacement platform (Thorlabs RB13M), the MCF and the tube were aligned. This alignment allowed displacing the fiber forward in micrometric steps until 0.5 mm of the loose end of the MCF were inserted in the ceramic tube. Thanks to the narrow difference between the diameter of the MCF and that of the ceramic tube, the MCF fitted tightly in it, avoiding any slack of the fiber during measurements.

Then, the ceramic tube was fixed to a precision translation stage (Thorlabs LTS150) in order to displace it vertically upwards and downwards with high precision, as this stage has, according to the manufacturer, a minimum achievable incremental movement of 0.1 μm. In this way, when the tube was displaced up and down, the fiber was bent and a triangle created, whose sides were the actual position of the fiber, its initial horizontal position ($L_f$) and the vertical displacement $d$, as indicated in Fig. 4a. This mechanism to apply bending to the fiber is similar to that used in[15,19,21,31]. According to Fig. 4a, the degrees of the bending angle are calculated as $\sin\theta = d/L_f$. According to it, as each step of $d$ upwards or downwards was of 100 μm, it meant that the applied bending was of 0.716° at each step until reaching 3.583°. This process was repeated every 30° of rotation of the MCF in the direction indicated in Fig. 4b until completing a whole rotation around its axis five times.

To interrogate the sensor, the setup shown in Fig. 4a was implemented. The light to the MCF was launched, via a 50:50 fiber optic coupler (FOC, Thorlabs TW1550R5A2), from a superluminescent diode (Safibra), which is a broadband light source centered at 1550 nm. The reflected light was split into two paths with another 50:50 FOC (Thorlabs TW1550R5A2). A portion of the reflected light was analyzed with a spectrometer (I-MON-512 High Speed, Ibsen Photonics) that allowed us tracking the wavelength shift with picometer accuracy. The other portion of light was detected with an InGaAs photodetector (Thorlabs PDA30B2). These components were selected for being off-the-shelf equipment and in order to make the interrogation setup cost-effective. However, the use of the devices reported in[32,33] as light source, and the device in[34] as photodetector could represent an alternative to our interrogation system due to their higher specifications, as they are aimed at the development of ultrafast photonic devices[35,36].

The comparison of the gathered spectra and their corresponding measurement of wavelength shift and light power variation of four cases in which the MCF has opposite core orientations against the same bending plane is shown in Figs. 5, 6, 7 and 8. Note that for the same applied bending direction and angle (in degrees), the changes in the spectra are fast and go from mainly light power variations to mainly wavelength shift progressively in clockwise direction of rotation. For the case in Fig. 5, it can be assumed that only light power variations take place, for the ones in Fig. 8 only wavelength shifts, and in each of the intermediate positions shown in Figs. 6 and 7, a specific linear combination of wavelength shift and light power change can be observed. For consecutive measured points (from 90° to 180° with blue spectra, and from 270° to 360° with red spectra), both variables have opposite trends: as one increases, the other decreases consecutively and vice versa. Moreover, in the upper row of each figure, it can be noticed that the evolution of the spectra and the measurement of both variables is practically identical but of opposite sign to that in the lower row, which matches with the opposite orientation of the cores in each figure. Thus, from this particular behavior, it can be concluded that by combining the measurement of both variables (wavelength shift and light power variations), any bending direction and amplitude can be measured, as unique trends and sensitivities at each core orientation of the MCF can be acknowledged. From the analysis of such spectral behavior, it is possible to detect accurately any bending direction in 360° and amplitudes for bending angles up to 3.583°. Moreover, such results suggest that an exhaustive alignment between cores and bending direction is not needed to devise such MCF-based vector bending sensor.

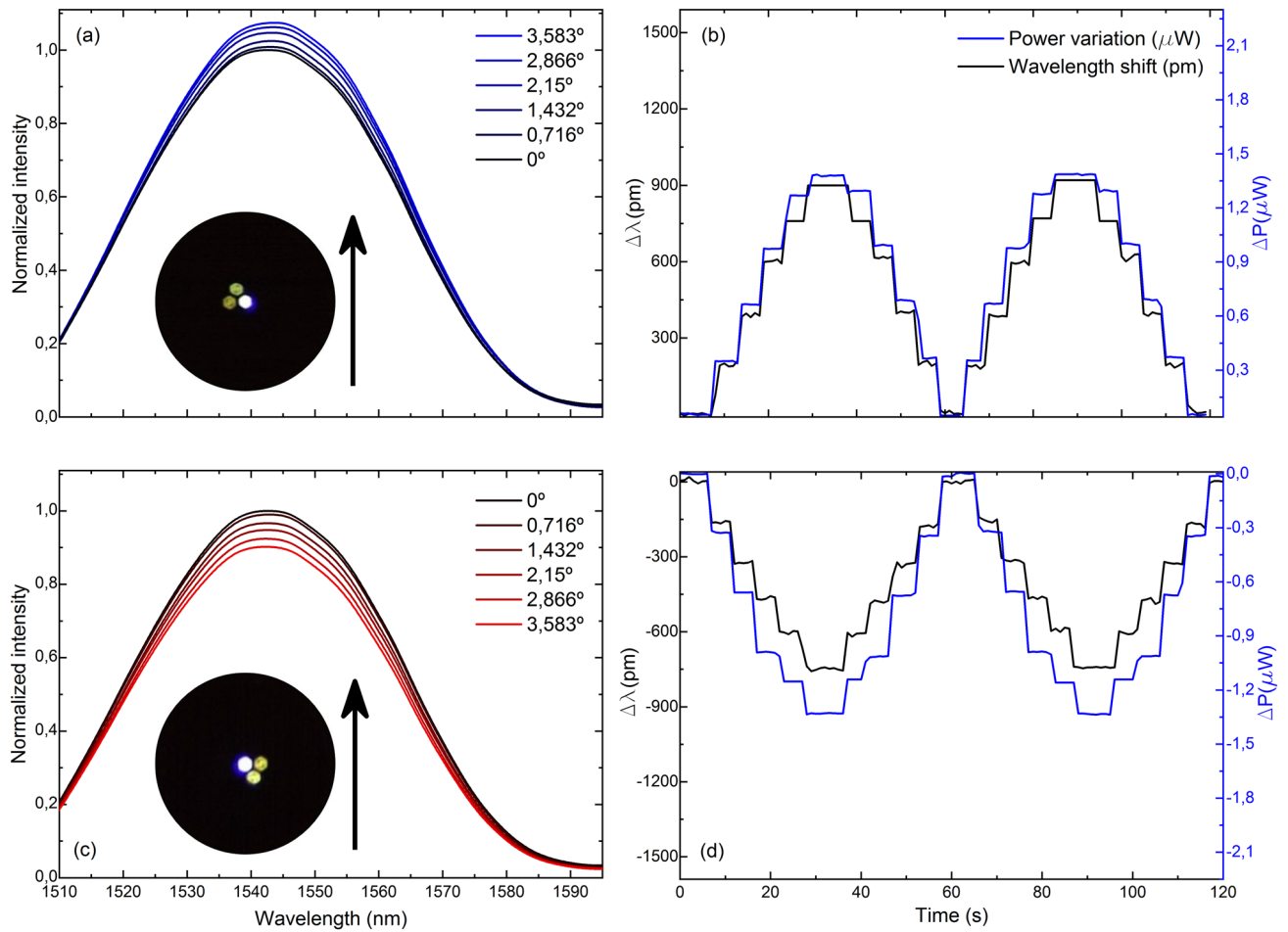

**Figure 6.** Reflection spectra and two cycles of the time evolution of the measured variables when the fabricated device is bent in the direction indicated by the black arrow and the MCF is rotated (**a**, **b**) 120° and (**c**, **d**) 300° with respect to the initial 0° position, respectively.

Regarding the temperature dependence of the proposed device, when MCFs as the one used in this work are exposed to temperature changes, proportional wavelength shifts are noticeable in the spectrum. Thus, if an SMF–MCF structure as the one shown in this work were bent and subjected to temperature changes simultaneously, the measured wavelength shift would be caused by both parameters at the same time. However, the response time of the shift caused by each parameter is different, which makes them easy to identify and differentiate. On the one hand, the response time is very fast for bending, the wavelength shift takes place as soon as the MCF is bent as a consequence of the instantaneous change in the effective refractive indexes responsible for the propagation characteristics. On the other hand, the response time is much slower when temperature changes happen. It takes some time since temperature is applied until its effects are noticeable in the spectrum, as the change in temperature has to reach all the cores of the MCF uniformly to have a stable signal. This effect is even slower if the fiber is protected or inserted in a capillary or a tube, as in our case, as the process of the heat transfer from the environment to the fiber is slowed down[37]. This slowdown does not happen for bending measurements regardless if it is inserted in a capillary or not, as it always takes place as soon as the fiber is bent. Thus, both shifts could be identified and distinguished by their response times, as fast wavelength shifts (high frequency components of the signal) are related to bending and slow wavelength shifts (low frequency components of the signal) are related to temperature. Regarding the power measurements, they are not affected by temperature variations, as the latter only causes wavelength shift. Additionally, power measurements are always normalized with respect to the power measurement when the fiber is straight, regardless of the temperature.

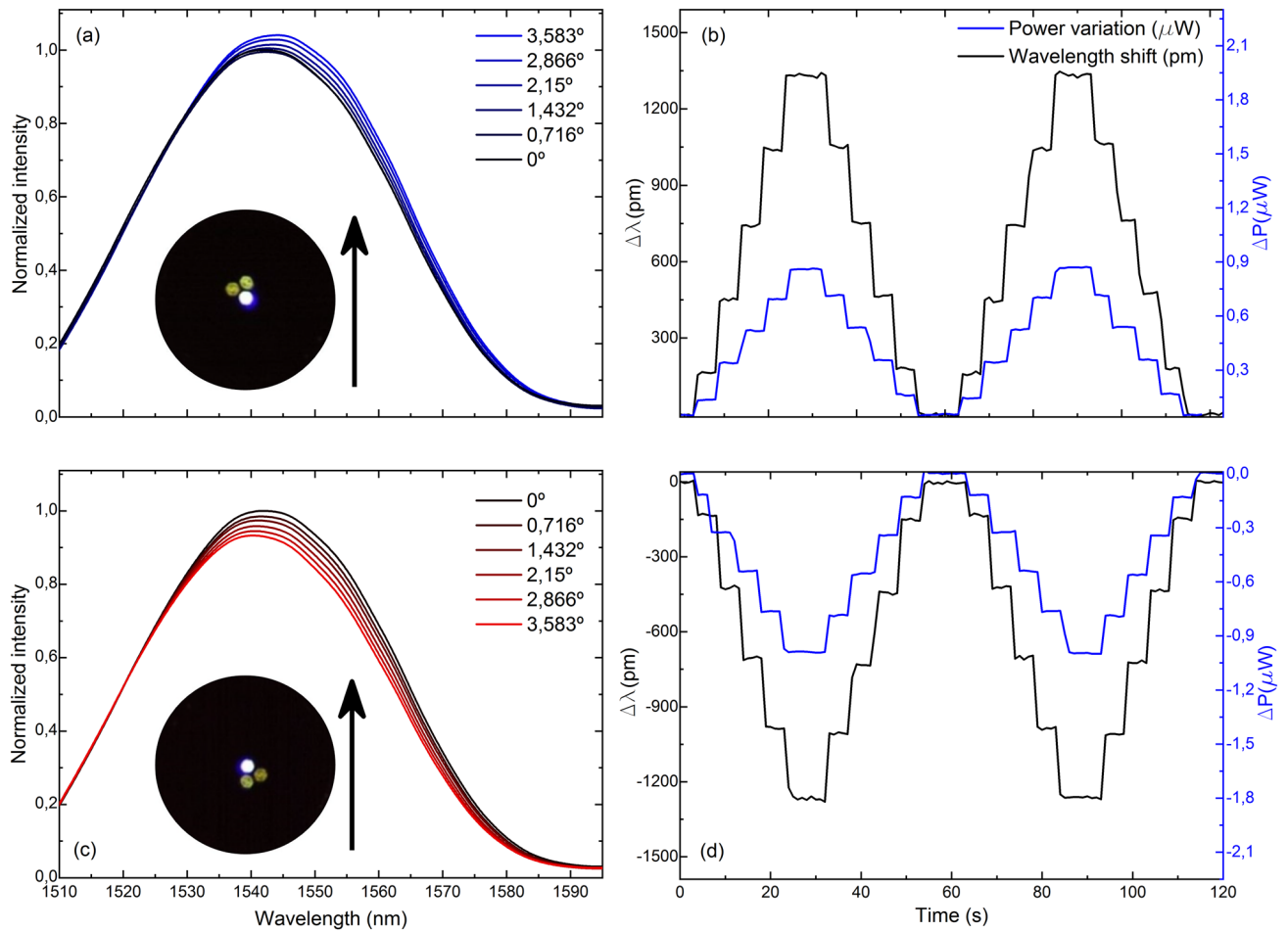

**Figure 7.** Reflection spectra and two cycles of the time evolution of the measured variables when the fabricated device is bent in the direction indicated by the black arrow and the MCF is rotated (**a**, **b**) 150° and (**c**, **d**) 330° with respect to the initial 0° position, respectively.

The bending sensitivities in terms of wavelength shift and light power variation and their respective standard deviation at the measured MCF orientations with respect to the applied bending direction are summarized in Fig. 9. The maximum sensitivities and their respective standard deviations were found to be 506.72 ± 5.50 pm/° and 587.50 ± 11.08 nW/°, for wavelength shift and light power variation measurements, respectively. These values represent an uncertainty in the measurement of 0.01° for wavelength shift and 0.018° for power variation. For the case of wavelength shift measurements, positive values indicate a shift to longer wavelengths compared to the position of the wavelength at which the maxima in the spectrum takes place when the MCF is straight, whereas negative values indicate a shift to shorter wavelengths. Similar explanation is valid for light power measurements: positive values indicate that the detected power is higher compared to the one received when the device is straight, and lower for negative values. The combination of the trend and sensitivity of each parameter is unique at each point, which leads to an unambiguous detection of the bending direction and amplitude. It can be noticed that both curves are 90° out of phase between them as well, as expected from the results in Figs. 5, 6, 7 and 8.

Figure 10 is the normalized polar representation of the results in Fig. 9, in absolute value. When the shift in the reflection spectrum is maximum, the light power change is minimal, and vice versa. Moreover, an asymmetrical behavior can be noticed as well within the same parameter in opposite points of the circumference, as for example for the normalized sensitivity in terms of wavelength shift at 0° and 180°, or in terms of light power variation at 90° and 270°. Each opposed couple of points in the polar representation corresponds to the MCF positions in which the outer cores have opposite orientations against the same applied bending direction, as shown in Figs. 5, 6, 7 and 8. In such situations, similar but not identical levels of stress among cores are induced in each case, which affects the normalized power coupling condition and explains the fact that the absolute value of the sensitivities in opposite points of the circumference are alike but not the same.

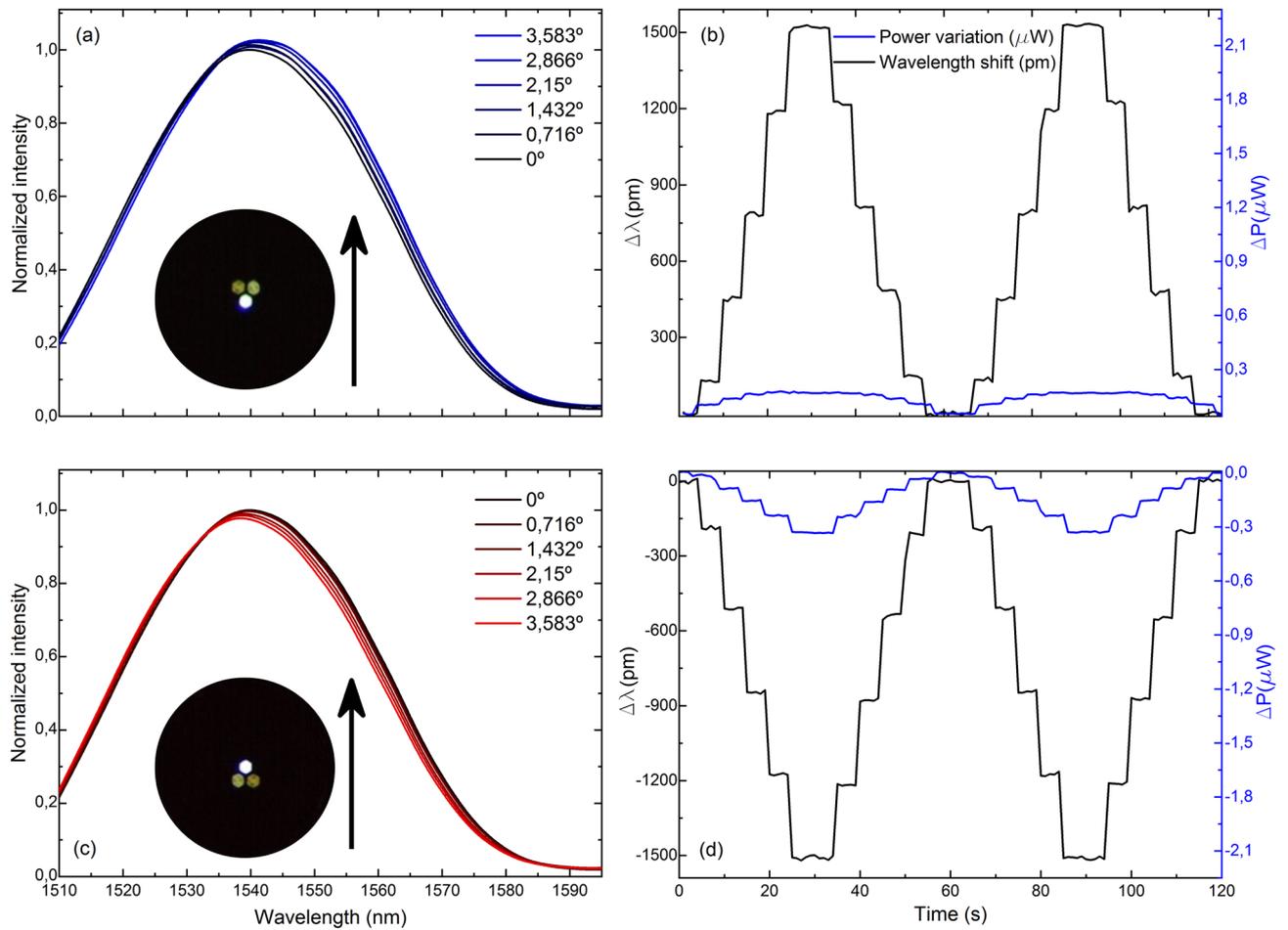

**Figure 8.** Reflection spectra and two cycles of the time evolution of the measured variables when the fabricated device is bent in the direction indicated by the black arrow and the MCF is rotated (**a**, **b**) 180° and (**c**, **d**) 360° with respect to the initial 0° position, respectively.

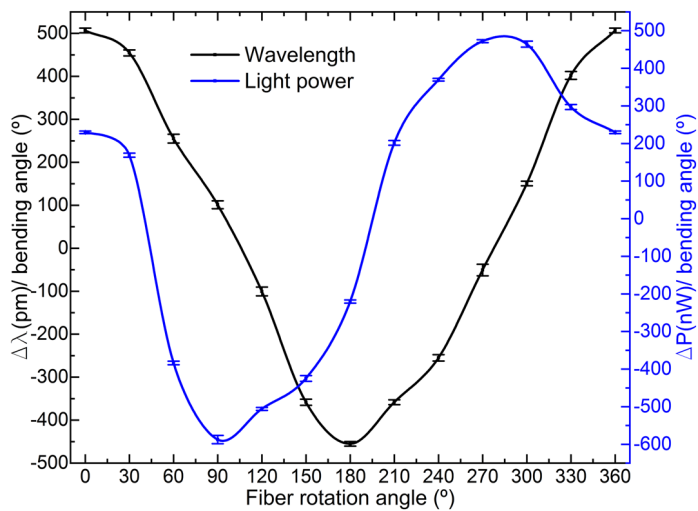

**Figure 9.** Wavelength shift and light power variation sensitivities and their respective standard deviations at each fiber position. In all cases identical bending direction and amplitude was applied. The combination of both parameters makes the response of the sensor unique at each point.

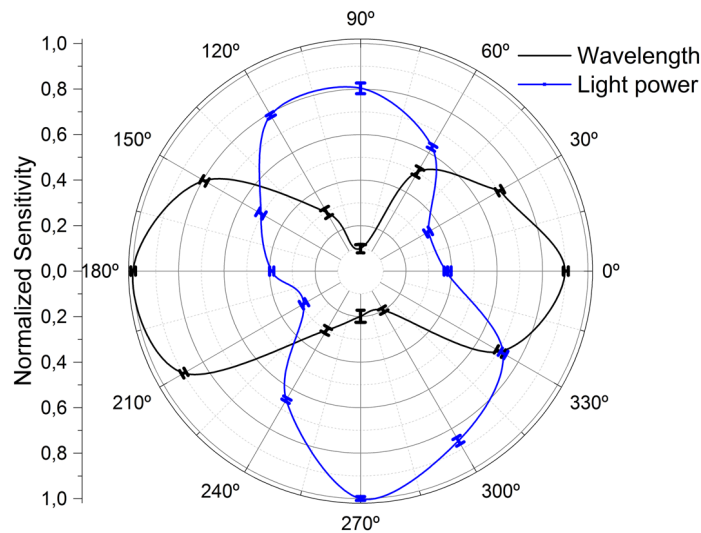

**Figure 10.** Polar representation of the normalized bending sensitivities and standard deviations in absolute value.

Therefore, we believe the bending sensor proposed here has potential to be deployed in real-world environments with an adequate packaging, as it was demonstrated that it has some valuable advantages as the capability of measuring bending direction and amplitude accurately in 360° and ease of interrogation. In addition, any alignment between the MCF cores with respect to the direction of bending is not necessary, which may simplify substantially the sensor packaging and its deployment. Moreover, it was demonstrated that our sensor can measure bending angles below 1° as well as large bending angles of up to 3.583°; hence, it can have multiple applications.

Some suggestions to make this device more appealing for real-world environments include changes in the interrogation setup to make it more cost-effective without affecting its performance. For example, a narrower light source with a flatter spectrum and a reference photodetector can be used to avoid the effect of any fluctuation of the emitted light source.

## Conclusions

In conclusion, we have reported on an omnidirectional vector bending sensor based on a short segment (8 mm) of an asymmetric coupled-core MCF spliced to a conventional SMF. By combining simultaneous wavelength shift and light power variation measurements, our sensor is capable of discerning bending direction and amplitude accurately in 360°. Other advantages of the sensor reported here include compactness, manufacturing simplicity, versatility (it is capable to operate for small and large bending angles, from below 1° up to 3.583°) and ease of implementation (no need of any specific alignment to operate and simple interrogation system). Thus, we believe the device is an appealing solution for the applications in which such requirements are needed.

### References
 1. Leng, J., Asundi, A. J. S. & Physical, A. A. Structural health monitoring of smart composite materials by using EFPI and FBG sensors. *Sens. Actuator A Phys.* **103**, 330–340 (2003).
 2. Gomez, J. *et al.* Comparing polymer optical fiber, fiber Bragg grating, and traditional strain gauge for aircraft structural health monitoring. *Appl. Opt.* **48**, 1436–1443 (2009).
 3. Yang, X. *et al.* Textile fiber optic microbend sensor used for heartbeat and respiration monitoring. *IEEE Sens. J.* **15**, 757–761 (2014).
 4. Leal-Junior, A. G., Frizera-Neto, A., Pontes, M. J. & Botelho, T. R. Hysteresis compensation technique applied to polymer optical fiber curvature sensor for lower limb exoskeletons. *Meas. Sci. Technol.* **28**, 125103 (2017).
 5. Martinez, A., Lai, Y., Dubov, M., Khrushchev, I. & Bennion, I. J. E. L. Vector bending sensors based on fibre Bragg gratings inscribed by infrared femtosecond laser. *Electron. Lett.* **41**, 472–474 (2005).
 6. Yong, Z., Zhan, C., Lee, J., Yin, S. & Ruffin, P. J. Multiple parameter vector bending and high-temperature sensors based on asymmetric multimode fiber Bragg gratings inscribed by an infrared femtosecond laser. *Opt. Lett.* **31**, 1794–1796 (2006).
 7. Shao, L.-Y., Xiong, L., Chen, C., Laronche, A. & Albert, J. J. Directional bend sensor based on re-grown tilted fiber Bragg grating. *J. Lightwave Technol.* **28**, 2681–2687 (2010).
 8. Chen, X. *et al.* Bragg grating in a polymer optical fibre for strain, bend and temperature sensing. *Meas. Sci. Technol.* **21**, 094005 (2010).
 9. Zhou, Q. *et al.* Bending vector sensor based on a sector-shaped long-period grating. *IEEE Photon. Technol. Lett.* **27**, 713–716 (2015).
10. Li, Y.-P. *et al.* Bending vector sensor based on a pair of opposite tilted long-period fiber gratings. *IEEE Photon. Technol. Lett.* **29**, 224–227 (2016).
11. Zhao, D. *et al.* Implementation of vectorial bend sensors using long-period gratings UV-inscribed in special shape fibres. *Meas. Sci. Technol.* **15**, 1647–1650. https://doi.org/10.1088/0957-0233/15/8/037 (2004).
12. Zhao, D. *et al.* Bend sensors with direction recognition based on long-period gratings written in D-shaped fiber. *Appl. Opt.* **43**, 5425–5428 (2004).

13. Zhang, S., Zhang, W., Gao, S., Geng, P. & Xue, X. Fiber-optic bending vector sensor based on Mach–Zehnder interferometer exploiting lateral-offset and up-taper. *Opt. Lett.* **37**, 4480–4482. https://doi.org/10.1364/OL.37.004480 (2012).
14. Zhang, L. *et al.* A fiber bending vector sensor based on M-Z interferometer exploiting two hump-shaped tapers. *IEEE Photon. Technol. Lett.* **27**, 1240–1243. https://doi.org/10.1109/LPT.2015.2415558 (2015).
15. Villatoro, J., Minkovich, V. P. & Zubia, J. J. Photonic crystal fiber interferometric vector bending sensor. *Opt. Lett.* **40**, 3113–3116 (2015).
16. Saffari, P. *et al.* Long period grating in multicore optical fiber: An ultra-sensitive vector bending sensor for low curvatures. *Opt. Lett.* **39**, 3508–3511 (2014).
17. Barrera, D., Gasulla, I. & Sales, S. Multipoint two-dimensional curvature optical fiber sensor based on a nontwisted homogeneous four-core fiber. *J. Lightwave Technol.* **33**, 2445–2450. https://doi.org/10.1109/JLT.2014.2366556 (2015).
18. Salceda-Delgado, G. *et al.* Compact fiber-optic curvature sensor based on super-mode interference in a seven-core fiber. *Opt. Lett.* **40**, 1468–1471. https://doi.org/10.1364/OL.40.001468 (2015).
19. Villatoro, J. *et al.* Ultrasensitive vector bending sensor based on multicore optical fiber. *Opt. Lett.* **41**, 832–835. https://doi.org/10.1364/OL.41.000832 (2016).
20. Yin, G., Zhang, F., Xu, B., He, J. & Wang, Y. Intensity-modulated bend sensor by using a twin core fiber: Theoretical and experimental studies. *Opt. Express* **28**, 14850–14858. https://doi.org/10.1364/OE.390054 (2020).
21. Arrizabalaga, O. *et al.* High-performance vector bending and orientation distinguishing curvature sensor based on asymmetric coupled multi-core fibre. *Sci. Rep.* **10**, 14058. https://doi.org/10.1038/s41598-020-70999-8 (2020).
22. Huang, W.-P. Coupled-mode theory for optical waveguides: An overview. *J. Opt. Soc. Am. A* **11**, 983 (1994).
23. Hudgings, J., Molter, L. & Dutta, M. Design and modeling of passive optical switches and power dividers using non-planar coupled fiber arrays. *IEEE J. Quant. Electron.* **36**, 1438–1444 (2000).
24. Snyder, A. W. Coupled-mode theory for optical fibers. *J. Opt. Soc. Am.* **62**, 1267–1277. https://doi.org/10.1364/JOSA.62.001267 (1972).
25. Xia, C., Bai, N., Ozdur, I., Zhou, X. & Li, G. Supermodes for optical transmission. *Opt. Express* **19**, 16653–16664. https://doi.org/10.1364/OE.19.016653 (2011).
26. Xia, C. *et al.* Supermodes in coupled multi-core waveguide structures. *IEEE J. Sel. Top. Quant. Electron.* **22**, 196–207 (2016).
27. Heiblum, M. & Harris, J. Analysis of curved optical waveguides by conformal transformation. *IEEE J. Quant. Electron.* **11**, 75–83. https://doi.org/10.1109/JQE.1975.1068563 (1975).
28. Schermer, R. T. Mode scalability in bent optical fibers. *Opt. Express* **15**, 15674–15701. https://doi.org/10.1364/OE.15.015674 (2007).
29. Schermer, R. T. & Cole, J. H. Improved bend loss formula verified for optical fiber by simulation and experiment. *IEEE J. Quant. Electron.* **43**, 899–909. https://doi.org/10.1109/JQE.2007.903364 (2007).
30. Amorebieta, J. *et al.* Highly sensitive multicore fiber accelerometer for low frequency vibration sensing. *Sci. Rep.* **10**, 16180. https://doi.org/10.1038/s41598-020-73178-x (2020).
31. Qu, H., Yan, G. F. & Skorobogatiy, M. Interferometric fiber-optic bending/nano-displacement sensor using plastic dual-core fiber. *Opt. Lett.* **39**, 4835–4838. https://doi.org/10.1364/OL.39.004835 (2014).
32. Guo, Y.-X., Li, X.-H., Guo, P.-L. & Zheng, H.-R. Supercontinuum generation in an Er-doped figure-eight passively mode-locked fiber laser. *Opt. Express* **26**, 9893–9900. https://doi.org/10.1364/OE.26.009893 (2018).
33. Feng, J., Li, X., Zhu, G. & Wang, Q. J. Emerging high-performance SnS/CdS nanoflower heterojunction for ultrafast photonics. *ACS Appl. Mater. Interfaces.* **12**, 43098–43105. https://doi.org/10.1021/acsami.0c12907 (2020).
34. Liu, J.-S. *et al.* SnSe2 nanosheets for subpicosecond harmonic mode-locked pulse generation. *Small* **15**, 1902811. https://doi.org/10.1002/smll.201902811 (2019).
35. Wang, L. *et al.* Few-layer mxene Ti3C2Tx (T = F, O, Or OH) for robust pulse generation in a compact Er-doped fiber laser. *ChemNanoMat* **5**, 1233–1238. https://doi.org/10.1002/cnma.201900309 (2019).
36. Feng, J. *et al.* SnS Nanosheets for 105th Harmonic Soliton Molecule Generation. *Ann. Phys.* **531**, 1900273. https://doi.org/10.1002/andp.201900273 (2019).
37. Amorebieta, J. *et al.* Packaged multi-core fiber interferometer for high-temperature sensing. *J. Lightwave Technol.* **37**, 2328–2334. https://doi.org/10.1109/JLT.2019.2903595 (2019).

### Acknowledgements
This work was funded in part by the Fondo Europeo de Desarrollo Regional (FEDER), in part by the Ministerio de Ciencia, Innovación y Universidades—under projects RTC2019-007194-4, RTI2018-0944669-B-C31 and PGC2018-101997-B—and in part by the Gobierno Vasco/Eusko Jaurlaritza IT933-16, ELKARTEK KK-2019/00101 (µ4Indust), and ELKARTEK KK-2019/00051 (SMARTRESNAK). The work of Josu Amorebieta is funded by a PhD fellowship from the University of the Basque Country UPV/EHU.

### Author contributions
The original draft of the paper was written by J.A. and A.O.-G., and reviewed by G.D., R.F. and J.V. J.A. collaborated in the theoretical approach, designed and performed the experiments, and processed and analyzed data. A.O.-G. did the simulations and the theoretical approach, J.V. conceived the device and supervised the experiments. A.S., E.A.-L. and R.A.C. conceived and fabricated the MCF. Project resources were provided by J.Z. All authors discussed the experimental data, revised and approved the manuscript. J.A. wrote the final version with inputs of all the authors.

### Competing interests
The authors declare no competing interests.

### Additional information
**Correspondence** and requests for materials should be addressed to J.A.

**Reprints and permissions information** is available at www.nature.com/reprints.

**Publisher's note** Springer Nature remains neutral with regard to jurisdictional claims in published maps and institutional affiliations.